\documentclass[pss]{wiley2sp}
\usepackage{amsmath}

\newcommand{\yfg}{YFe$_2$Ge$_2$}

\begin{document}
\title{Fermi liquid breakdown and evidence for superconductivity in \yfg}
\titlerunning{Fermi liquid breakdown in \yfg}
\author{Y. Zou\textsuperscript{\textsf{\bfseries 1}},
  Z. Feng\textsuperscript{\textsf{\bfseries 2}},
  P. W. Logg\textsuperscript{\textsf{\bfseries 1}},
  J. Chen\textsuperscript{\textsf{\bfseries 1}},
  G. Lampronti\textsuperscript{\textsf{\bfseries 3}},
  F. M. Grosche\textsuperscript{\Ast\textsf{\bfseries 1}}}

\authorrunning{Y. Zou et al.}
\mail{e-mail
  \textsf{fmg12@cam.ac.uk}, Phone:
  +44-1223-337392}

\institute{%
  \textsuperscript{1}\,Cavendish Laboratory, University of Cambridge, J J Thomson Ave, Cambridge CB3 0HE, UK\\
  \textsuperscript{2}\,London Centre of Nanotechnology, University College London, Gordon Street, London WC1H 0AH, UK\\
  \textsuperscript{3}\,Dept. of Earth Sciences, University of Cambridge, Downing
  Street, Cambridge CB2 3EQ}

\received{XXXX, revised XXXX, accepted XXXX} 
\published{XXXX} 

\keywords{Non-Fermi liquid, superconductivity, transition metal compounds}


\abstract{
\abstcol{
  In the d-electron system YFe$_2$Ge$_2$, an unusually high and temperature
  dependent Sommerfeld ratio of the specific heat capacity $C/T \sim
  100~\rm{mJ/(molK^2)}$ and an anomalous power law temperature
  dependence of the electrical resistivity $\rho \simeq \rho_0 +
  AT^{3/2}$ signal Fermi liquid breakdown, probably connected to a
  close-by quantum critical point.}{Full resistive transitions,
  accompanied by DC diamagnetic screening fractions of up to 80\%
  suggest that pure samples of YFe$_2$Ge$_2$\ superconduct below
  $1.8~\rm{K}$.}}

\maketitle

The threshold of magnetism in transition metal compounds is frequently
associated with anomalous low temperature properties, such as the
robust $T^{3/2}$ power-law resistivity observed in MnSi
\cite{pfleiderer01}, ZrZn$_2$ \cite{takashima07,smith08} and NbFe$_2$
\cite{brando08}. Quantum critical phenomena associated with incipient
antiferromagnetic or spin density wave order remain comparatively
underexplored in this material class. The close association of
superconductivity with the border of antiferromagnetism in a large
number of iron pnictide and chalcogenide compounds further motivates
the search for suitable candidate materials among transition metal intermetallics.

The (Y/Lu)Fe$_2$Ge$_2$ system offers such an opportunity to study a
spin density wave quantum phase transition in a transition metal
composition series. LuFe$_2$Ge$_2$ crystallizes in the ThCr$_2$Si$_2$ 
structure ($I4/mmm$) and exhibits spin density wave order below $T_N =
9~\rm K$ \cite{avila04,ferstl06}. Electron counting places Fe in
(Y/Lu)Fe$_2$Ge$_2$ at the same valence as in the isostructural
superconductors (K, Rb, Cs)Fe$_2$As$_2$, but the magnetic order in LuFe$_2$Ge$_2$
differs from that of the iron arsenides: the Fe moments align
ferromagnetically within the basal plane and couple
antiferromagnetically to their neighbours along the crystallographic
$c$ direction, corresponding to an ordering wavevector ${\bf Q} = (
0~0~1)$ \cite{fujiwara07}. Applied hydrostatic pressure raises
$T_N$\cite{fujiwara07}. Partial substitution of Lu by Y expands the
unit cell and suppresses $T_N$ \cite{Ran11} at a critical composition
of Lu$_{0.8}$Y$_{0.2}$Fe$_2$Ge$_2$.

Here, we concentrate on the end member of the series, YFe$_2$Ge$_2$, which is
paramagnetic at ambient pressure and displays an unusually high
Sommerfeld coefficient of the specific heat capacity $C/T \sim 100 ~\rm
{mJ/mol K^2}$  \cite{avila04}. Our low temperature measurements 
reveal hallmarks of Fermi liquid breakdown, such as an anomalous $T^{3/2}$
power law form of the electrical resistivity and a strongly temperature
dependent $C/T$ at low $T$, as well as full resistive
superconducting transitions and large diamagnetic screening fractions
consistent with bulk superconductivity below an onset $T_c\simeq 1.8~\rm K$.

\begin{figure}[b]
  \includegraphics[width=\columnwidth]{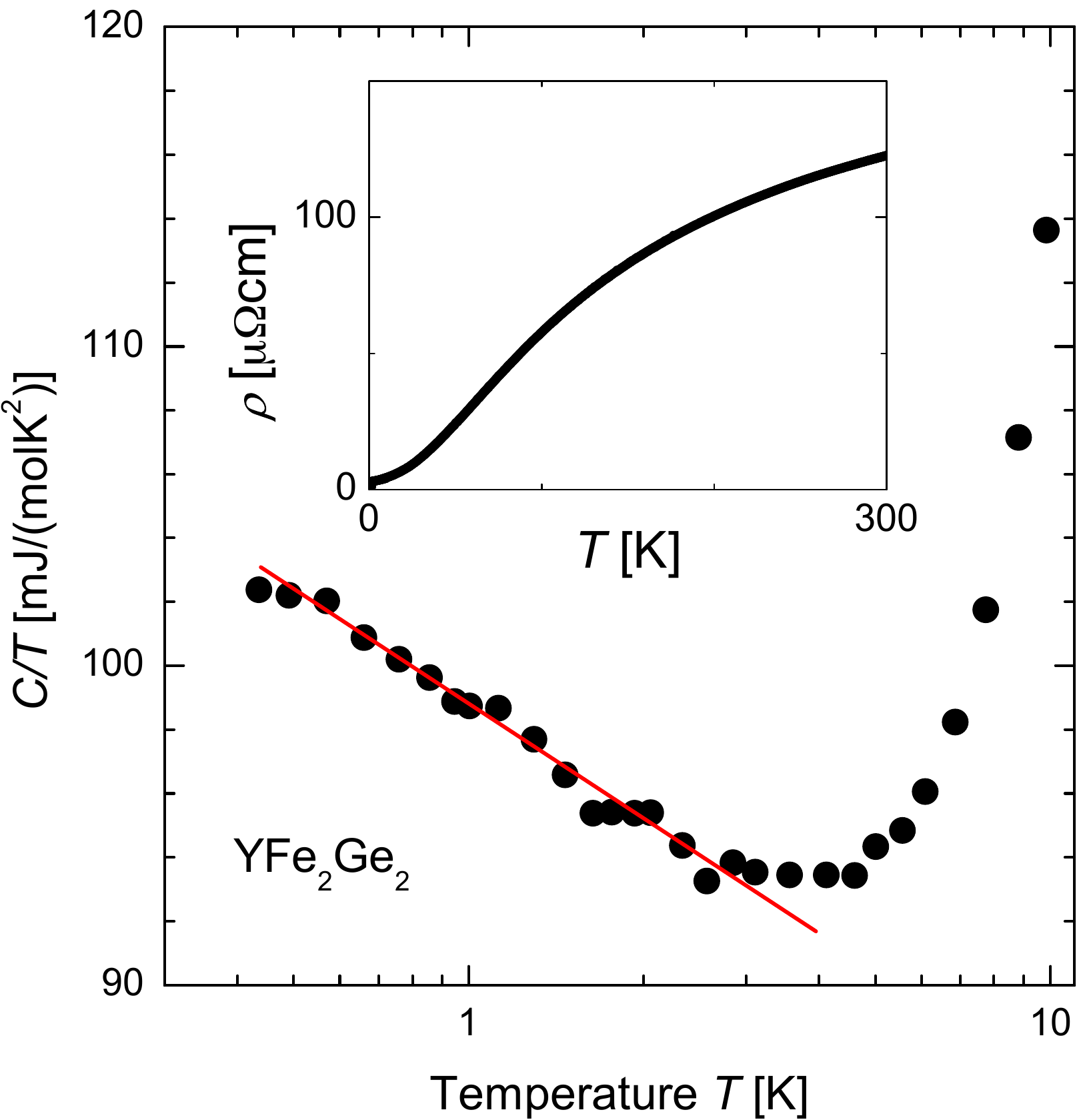}
  \caption{\label{fig:heat} Sommerfeld ratio of the specific heat
    capacity, $C/T$, vs. temperature $T$ in polycrystalline
    YFe$_2$Ge$_2$. The Sommerfeld ratio rises with decreasing temperature
    $T<4~\rm{K}$ and reaches values exceeding
    $100\,\mathrm{mJ/(molK^2)}$. Higher resolution, lower temperature
    data will be required to distinguish between a logarithmic
    $T$-dependence and the $\gamma_0 -a\sqrt{T}$ form predicted near a
    3D spin density wave quantum critical point. (inset) Electrical
    resistivity of polycrystalline YFe$_2$Ge$_2$ versus temperature.}
\end{figure}

Single crystals of YFe$_2$Ge$_2$ were grown from tin flux following
published methods \cite{avila04}, and polycrystals were obtained
by radio frequency melting of the elements (Y 3N, Fe 4N, Ge 6N) on a
water-cooled copper boat, followed by annealing in vacuum at
$800^\circ~\rm C$ for 7 days. The residual resistivity ratios
($\mathrm{RRR}=\rho(300\,\mathrm{K})/\rho(2\,\mathrm{K})$) obtained
from flux growth were typically about 10, whereas annealed
polycrystals reached resistivity ratios of up to 50. The electrical
resistivity and the magnetic susceptibility were measured in an
adiabatic demagnetisation refrigerator to below $0.1\,\mathrm{K}$, and
the specific heat capacity was measured in a Quantum Design Physical Properties
Measurement System with a $^3$He insert to below
$0.4\,\mathrm{K}$. The magnetisation data was acquired in a Cryogenic
SQUID magnetometer with a $^3$He insert to below
$0.3\,\mathrm{K}$. X-ray studies confirmed the quality and composition
of our samples, giving the lattice parameters $a=3.964(6) ~\rm \AA$,
$c=10.457(4)~\rm \AA$ and the conventional unit cell volume
$V=164.37(2)~\rm \AA^3$. Our samples were found to be at least $99\%$
phase pure, and the only impurity phase which could be identified in
some of the polycrystals is a ferromagnetic Fe/Ge alloy.

\begin{figure}[t]
\includegraphics[width=\columnwidth]{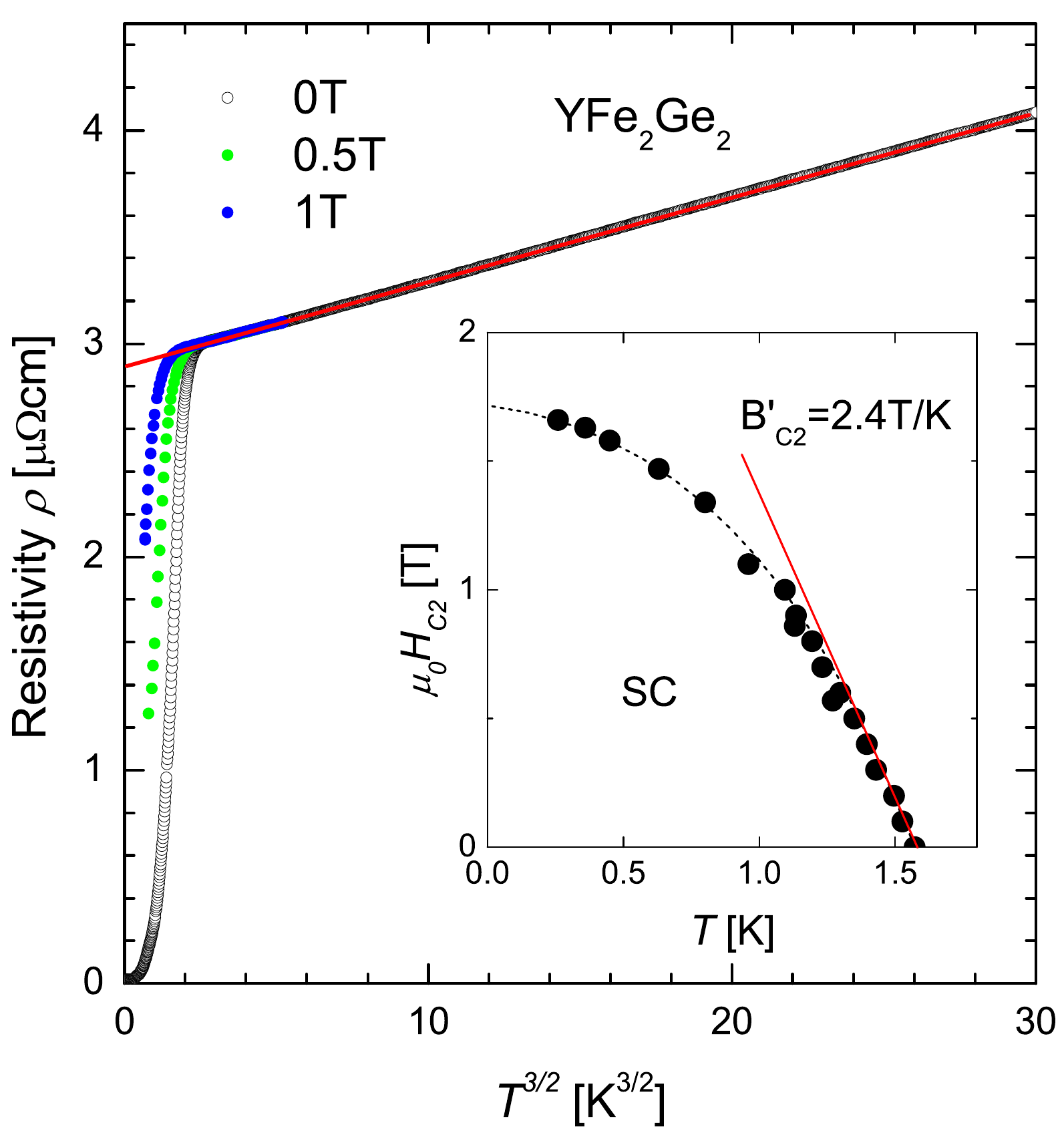}
\caption{\label{fig:res} Resistivity of polycrystalline YFe$_2$Ge$_2$
  with RRR=41, showing full resistive superconducting transitions at
  $T<1.8\,\mathrm{K}$ and a $T^{3/2}$ temperature dependence of the
  normal state resistivity. The superconducting transition is
  suppressed by moderate magnetic fields. (inset) Temperature
  dependence of the superconducting upper critical field, obtained
  from a partial transition criterion. The
  initial slope is $B'_{ c2} \simeq 2.4\,\mathrm{T/K}$ (inset) and the
  extrapolated zero-temperature critical field is
  $H_{c2}=1.7\,\mathrm{T}$.}
\end{figure}

At elevated temperatures $T>2~\rm K$, our findings are
consistent with those reported previously \cite{avila04}: The
magnetic susceptibility is small and weakly temperature dependent, the
resistivity is metallic (inset of Fig.~\ref{fig:heat}), and the
Sommerfeld coefficient of the specific heat capacity,
$C/T$ is surprisingly high, reaching values near $\simeq
100~\mathrm{mJ/molK^2}$ at $2~\rm K$ (Fig.~\ref{fig:heat}).

Extending the measurements to lower $T$ reveals a gradual further
increase of $C/T$ (Fig.~\ref{fig:heat}), and the electrical
resistivity displays an unusual power-law temperature dependence of
the form $\rho(T) \simeq \rho_0 + AT^{3/2}$ up to temperatures of the
order of $10~\rm K$ (Fig.~\ref{fig:res}). These findings suggest Fermi
liquid breakdown similar to that observed in other transition metal
compounds, such as MnSi, ZrZn$_2$ and NbFe$_2$
\cite{pfleiderer01,takashima07,smith08,brando08}. In contrast to the
latter, which are close to the threshold of ferromagnetism, the weak
$T$-dependence and comparatively small magnitude of the magnetic
susceptibility in YFe$_2$Ge$_2$ \cite{avila04} suggests a different
scenario.  Electronic structure calculations \cite{subedi14,singh14}
indicate that ferromagnetism and various antiferromagnetic states
compete in YFe$_2$Ge$_2$. However, they find a significant energy
advantage for ${\bf Q}=(0~0~1)$ order, which is indeed observed in
LuFe$_2$Ge$_2$. This would be consistent with the interpretation that
the low temperature properties of YFe$_2$Ge$_2$ are affected by a
nearby antiferromagnetic (or spin density wave) quantum critical
point. Proximity to a magnetic quantum critical point could also
explain one of the central puzzles in YFe$_2$Ge$_2$, namely the
10-fold enhancement of the heat capacity over the band structure value
of $\sim 10~{\rm mJ/(molK^2)}$ \cite{subedi14,singh14}.  However,
stoichiometric YFe$_2$Ge$_2$ is sufficiently far removed from the critical
composition found in the (Y/Lu)Fe$_2$Ge$_2$ composition series
\cite{Ran11} to leave open alternative possibilities.

\begin{figure}[b]
\includegraphics[width=\columnwidth]{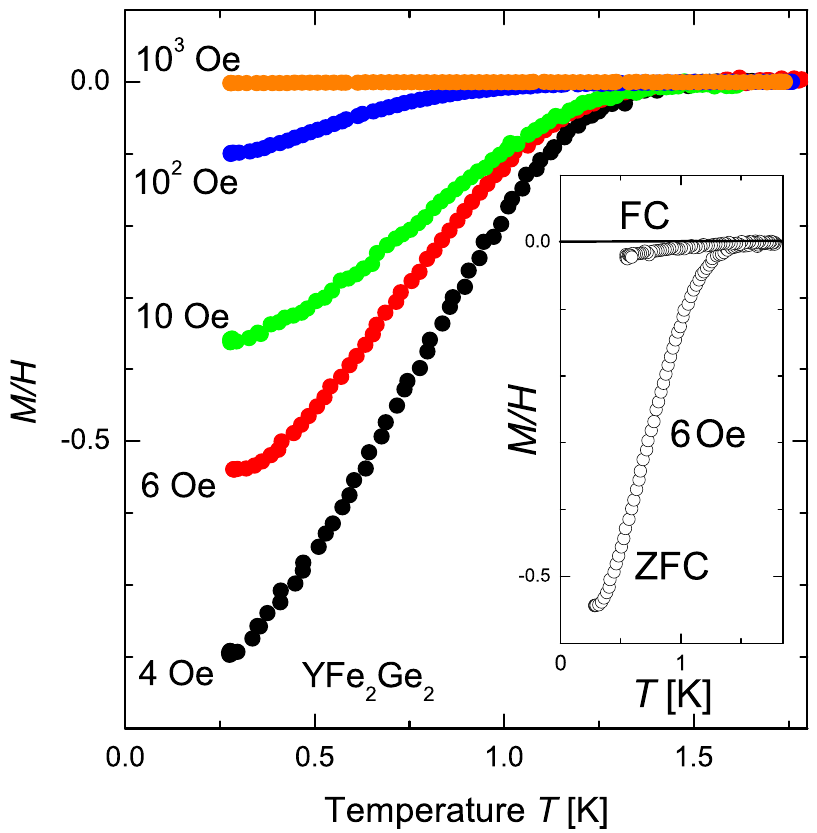}
\caption{\label{fig:mag} DC magnetisation data in
  YFe$_2$Ge$_2$. A strong diamagnetic signal is observed in zero-field
  cooled measurements below $1.5~\rm{K}$. The size
  and position of the diamagnetic signature shifts with increasing
  magnetic field, suggesting that the lower critical field is around
  $2~\rm{mT}$. (inset) Zero-field cooled and field cooled magnetic
  response at very low applied field.}
\end{figure}

Below $2~\rm{K}$, superconducting transitions are apparent in the
electrical resistivity (Fig.~\ref{fig:res}) as well as in the DC
magnetisation (Fig.~\ref{fig:mag}). While the onset of the resistive
transition is observed at $T_c ^{\rho} \simeq 1.8~\rm{K}$ ($50\%$
point at $\simeq 1.4 ~\rm K$), the onset of superconductivity is
apparent in DC magnetisation measurements only below $T_c^{\rm mag}
\simeq 1.5~\rm{K}$. This suggests that $T_c$
varies across the sample. For the sample shown, the superconducting
volume fraction extracted from the magnetisation measurement
(Fig.~\ref{fig:mag}) amounts to at least $80 \%$.

Comparing polycrystals with different residual resistance ratios, we
find a clear correlation between sample purity and both the size of
the superconducting jump and the value of $T_c$ observed in
resistivity measurements. Full superconducting transitions are
obtained in polycrystalline samples with $\rm{RRR} \simeq 30-50$,
while unannealed polycrystals with lower RRR as well as the flux-grown
single crystals with $\rm{RRR} \simeq 10$ only show partial
transitions.

Heat capacity measurements down to $400~\rm{mK}$ have not
revealed a clear anomaly associated with the superconducting
transition. This might result from an intrinsically interesting
mechanism, such as an anomalously reduced superconducting
gap. However, at this stage it could equally be attributed to the
broad nature of the transition, or it could suggest that despite the
large diamagnetic screening signal superconductivity is confined to
small fractions of the sample. A comprehensive study on high quality
flux-grown crystals \cite{kim14} also remarked on the absence of a
heat capacity anomaly near $T_c$, whereas a superconducting heat
capacity anomaly was reported by a third group \cite{strydom14}. This 
indicates that if alien phases can be ruled out, slight variations in
the stoichiometry might play a role in establishing bulk
superconductivity in YFe$_2$Ge$_2$.

The initial slope of the resistive upper critical field is
$|dB_{c2}/dT| \simeq 2.4\,\mathrm{T/K}$ (inset of Fig.~\ref{fig:res}).
This corresponds to an extrapolated clean-limit weak-coupling
orbital-limited critical field $B^{(o)}_{c2} \simeq 0.73~ T_c ~
|dB_{c2}/dT| \simeq 2.8\,\mathrm{T}$ \cite {Helfand66}. This value
significantly exceeds the observed critical field in the low
temperature limit of $\simeq 1.7 ~\rm T$, suggesting that the low
temperature critical field is Pauli limited. In the standard treatment
(e.g. \cite{tinkham04}), the extrapolated orbital-limited critical
field corresponds to a superconducting coherence length $\xi_0 =
\left(\Phi_0/(2\pi B^{(o)}_{c2}\right)^{1/2} \simeq 108 ~\rm \AA$,
where $\Phi_0 = h/(2e)$ is the quantum of flux.  Such a short
coherence length is roughly consistent with the enhanced quasiparticle
mass and consequently low Fermi velocity indicated by the high
Sommerfeld coefficient of the specific heat capacity: we estimate the
BCS coherence length from $\xi_{BCS} = (\hbar v_F )/(\pi \Delta)$
\cite{tinkham04,orlando79}, where $v_{F}$ is the Fermi velocity and
$\Delta$ is the superconducting gap, approximated as $1.76 {\rm k_B}
T_c$. Representing the electronic structure of YFe$_2$Ge$_2$ by one or
two spherical Fermi surface sheets with radius $k_F = 0.7
~\rm\AA^{-1}$ (corresponding to half-filled bands) in order to extract
$v_F$ from $\gamma$, we would expect $\xi_{BCS} \simeq 130 ~\rm \AA$
for a single sheet or $260~\rm \AA$ for two sheets, in rough agreement
with $\xi_0$ obtained above. The mean free path in our samples can
likewise be estimated (e.g. \cite{orlando79}) from $\ell \simeq 1200
~\rm \AA (\AA^{-2}/k_F^2)(\mu\Omega cm/\rho_0)$, where $\rho_0$ is the
residual resistivity, to be about $800~\rm \AA$ ($400~\rm \AA$) for
the highest quality samples, when one (two) Fermi surface sheets are
assumed. This indicates that $\ell$ indeed exceeds $\xi_0$, so that an
anisotropic order parameters is not ruled out by disorder scattering.




We thank C. Geibel, P. Niklowitz, S. Friedemann, M. Gamza and
G. G. Lonzarich for helpful discussions. This work was supported by
EPSRC UK and Trinity College Cambridge.

\bibliographystyle{pss}

\providecommand{\WileyBibTextsc}{}
\let\textsc\WileyBibTextsc
\providecommand{\othercit}{}
\providecommand{\jr}[1]{#1}
\providecommand{\etal}{~et~al.}

\end{document}